\documentclass[twocolumn,english,aps,pra,preprintnumbers]{revtex4}
\usepackage{lmodern}
\usepackage[T1]{fontenc}
\usepackage[latin9]{inputenc}
\usepackage{amsmath}
\usepackage{amssymb}
\usepackage{graphicx}

\makeatletter

\providecommand{\tabularnewline}{\\}

\@ifundefined{textcolor}{}
{%
 \definecolor{BLACK}{gray}{0}
 \definecolor{WHITE}{gray}{1}
 \definecolor{RED}{rgb}{1,0,0}
 \definecolor{GREEN}{rgb}{0,1,0}
 \definecolor{BLUE}{rgb}{0,0,1}
 \definecolor{CYAN}{cmyk}{1,0,0,0}
 \definecolor{MAGENTA}{cmyk}{0,1,0,0}
 \definecolor{YELLOW}{cmyk}{0,0,1,0}
}

\makeatother

\usepackage{babel}
\begin{document}

\title{Vibrational spectroscopy of polar molecules with superradiance}

\author{Guin-Dar Lin and Susanne F. Yelin}

\affiliation{ITAMP, Harvard-Smithsonian Center for Astrophysics and Harvard Physics
Department, Cambridge, Massachusetts 02138, USA}

\affiliation{Department of Physics, University of Connecticut, Storrs, Connecticut
06269, USA}
\begin{abstract}
We investigate cooperative phenomena and superradiance for vibrational
transitions in polar molecule spectroscopy when a high optical-depth
(OD) sample is studied. Such cooperativity comes from the build-up
of inter-particle coherence through dipole-dipole interactions and
leads to the speed-up of decay process. We compare our calculation
to recent work {[}Deiglmayr et al., Eur. Phys. J. D 65, 99 (2011){]}
and find very good agreement, suggesting that superradiant effects
need to be included in a wide variety of ultracold molecule setups
including vibrational and rotational states.
\end{abstract}
\maketitle

\section{Introduction}

Ultracold molecules have recently become an intensely explored area
due to their fascinating properties. Many proposals have been brought
up, for example, to improve the precision measurement of fundamental
constants \cite{flambaum2006pra,flambaum2007prl}, or test the fundamental
law of physics, \textit{e.g.}, \cite{hudson2002prl}. As expected,
the question of how to gain better access to internal degrees of a
molecule to help better manipulation becomes an important step. The
most commonly used tool of analysis in molecule systems is spectroscopy.
However, for molecules the resolution of complicated spectral lines
with necessary accuracy is crucial. Typically, the spectroscopic model
for internal degrees of a molecule has also been applied to an ensemble,
by assuming that the overall radiative behavior does not deviate dramatically
from the direct sum of individual responses of the ingredient molecules.
Even in the case of amplified spontaneous emission (ASE, also called
``superluminescence'') \cite{Malcuit1987,Bolda1995}, where the medium
nonlinearity plays an important role, the inter-particle coherence
in terms of genuine quantum manybody nature is usually overlooked.
Such quantum manybody coherence can modify the emission rate significantly,
resulting in ``superradiance'', which was first predicted by Dicke
in 1954 \cite{Dicke1953}. Since then plenty of efforts have been
made to investigate such phenomena both theoretically \cite{Arecchi1970,Lehmberg1970,Rehler1971,Stroud1972,Friedberg1974,Bonifacio1975a,Gross1982,Bolda1995}
and experimentally \cite{Feher1958,Skribanowitz1973,Vrehen1977,Gounand1979,Raimond1982,Goy1983,Moi1983,Malcuit1987}.
The interest in superradiance has been revived especially during recent
years due to the advances of controlling atomic, molecular, and other
quantum optical systems \cite{Inouye1999,Scheibner2007,Bohnet2012}.

Looking at the simplest model, Dicke considered a case where the ensemble
of excited two-level particles is confined in space to a size much
smaller than the transition wavelength. In the limiting case where
we are allowed to regard the whole ensemble as a point-like system,
the system can be described by a collective spin $J=N/2$ with $N$
the number of particles and the factor $\frac{1}{2}$ is for two-level
atoms analogous to spin-$\frac{1}{2}$ particles. As a result the
overall decay rate is now given by $\Gamma(J,M)=\gamma\langle JM|D^{+}D^{-}|JM\rangle$,
where $\gamma$ is the free-space spontaneous decay rate, the collective
raising operator $D^{+}=\sum_{i}|e\rangle_{i}\langle g|$ with particle
index $i$ and the lowering operator $D^{-}=(D^{+})^{\dagger}$; $M$
is the index of the Dicke ladder with $-J\le M\le J$. This factor
$\langle JM|D^{+}D^{-}|JM\rangle=(J-M)(J+M-1)$ quantifies the enhancement
of the decay rate, contributing to the origin of superradiance \cite{Dicke1953,Gross1982}.

Note that in the above discussion for Dicke superradiance, the sample
size is considered infinitesimally small compared to the transition
wavelength $\lambda$. In reality, the particle ensemble must have
a finite spread in space. Also, the distance-dependent dipole-dipole
interaction breaks the Dicke symmetry and causes considerable dephasing.
A quantity called cooperativity can be defined to characterize the
competition of these two length scales: $\mathcal{C}\equiv\mathcal{N}\lambda^{3}/(4\pi)$,
where $\mathcal{N}$ is the volume density of particles. As one can
expect, superradiance is more obvious when $\mathcal{C}\gg1$ but
getting suppressed when $\mathcal{C}$ decreases. However, according
to further analysis \cite{Arecchi1970,MacGillivray1976,Chen1999,Akkermans2008,Lin2012,lin2012adv6},
it is more accurate to characterize superradiance behavior by the
optical depth $\text{OD}\equiv\mathcal{N}\lambda^{2}d$ with sample
diameter $d$, consistent with the known fact that the superradiance
can be directional and hence depends on the geometry of the sample.
Recent experimental technology has allowed to realize high OD for
atomic and molecular ensembles. Therefore, cooperative phenomena such
as superradiance should be observable in such systems. Indeed, a study
on rubidium Rydberg atoms dealing with Rydberg state relaxation has
showed that the associated life time is considerably shortened given
$\text{OD}\approx10^{5}$ \cite{Farooqi2003,Wang2007}. Its decay
behavior agrees quite well with the theoretical calculation in our
previous work \cite{Wang2007}, suggesting the emergence of superradiance. 

In this paper, we want to discuss the same effects for polar molecules
specially for vibrational spectroscopy for which the parameter regime
is accessible by current technology. We have developed an effective
two-body formalism that accounts for the actual dipole-dipole interaction
for finite OD, for which the Dicke picture fails to be valid. We also
compare our calculation to a recent experiment on the vibrational
relaxation of LiCs molecules \cite{deiglmayr2011epjd}. Even though
in this experiment OD is rather small such that the superradiance
feature is not dominant, we observe a discrepancy between the measured
data and the results given by the traditional single-particle rate
equation approach including blackbody radiation. Such disagreement
is not found when superradiance physics is considered. We are then
able to conclude that the same cooperativity which leads to superradiance
for a higher OD still plays an important role here.

This paper is organized as follows: Sec. \ref{sec:dicke} discusses
the vibrational transitions in the Dicke-superradiance point of view,
addressing on the multilevel contribution. Various molecules are also
compared according to their vibration relevant parameters in order
for better demonstration of superradiance. Sec. \ref{sec:Formalism}
gives the effective master equation based on a mean-field approach,
where the two-body coherence is explicitly kept. Then we show our
calculation results in Sec. \ref{sec:Result}, where we also make
a comparison for the measured data with our theoretical prediction,
taking into account the photon and particle losses as well as the
thermal contribution due to blackbody radiation. Finally, Sec. \ref{sec:Conclusion}
summarizes this work and future outlook.

\section{Vibrational transitions in the Dicke limit\label{sec:dicke}}

\begin{table}
\begin{centering}
\begin{tabular}{|c|c|c|c|c|}
\hline 
 & $\mu$ (u) & $\omega_{e}$ (cm$^{-1}$) & $\wp$ (D) & OD ($\times10^{3}$)\tabularnewline
\hline 
\hline 
$^{7}$Li$^{39}$K & 5.95 & 207 & 3.4 & 1.2\tabularnewline
\hline 
$^{7}$Li$^{85}$Rb & 6.48 & 185 & 4.0 & 1.5\tabularnewline
\hline 
$^{7}$Li$^{133}$Cs & 6.66 & 167 & 5.5 \cite{aymar2005jcp,deiglmayr2011epjd} & 1.8\tabularnewline
\hline 
$^{23}$Na$^{39}$K & 14.5 & 124 & 2.8 \cite{aymar2005jcp} & 3.3\tabularnewline
\hline 
$^{23}$Na$^{85}$Rb & 18.1 & 107 & 3.1 & 4.4\tabularnewline
\hline 
$^{23}$Na$^{133}$Cs & 19.6 & 98 & 4.7 & 5.2\tabularnewline
\hline 
$^{39}$K$^{85}$Rb & 26.7 & 76 & 0.57 \cite{ni2008science} & 8.7\tabularnewline
\hline 
$^{39}$K$^{133}$Cs & 30.1 & 66 & 1.9 \cite{aymar2005jcp} & 12\tabularnewline
\hline 
$^{85}$Rb$^{133}$Cs & 51.8 & 49 & 1.3 \cite{aymar2005jcp} & 21\tabularnewline
\hline 
Rb (Rydberg) &  & 33 & $\sim10^{3}$ & 83\tabularnewline
\hline 
\end{tabular}
\par\end{centering}

\caption{\label{tab:polmol} Table of the reduced masses $\mu$, vibrational
spacing $\omega_{e}$, dipole moments $\wp$, and optical depths OD,
for various alkali diatomic molecules. To estimate OD , we assume
$\mathcal{N}=5\times10^{9}\text{cm}^{-3}$ and $d=100\mu\text{m}$.
Note that in \cite{deiglmayr2011epjd} the LiCs sample's OD$\approx500$
is rather small, with a lifetime $\sim20$ s. The corresponding parameters
($n=40\rightarrow39$) for rubidium Rydberg atoms from \cite{Wang2007}
are also listed for comparison. All data are from \cite{nistchemweb}
(and references therein) except for those otherwise specified. }
\end{table}

In order to see to what degree the superradiance effect must be considered,
here we first discuss the optical depth (OD) for different setup of
molecules. As it is expected that a larger OD makes superradiance
more obvious, a closer vibrational spacing is wanted and hence a larger
wavelength. Therefore those molecules with heavier reduced mass become
better candidates. On the other hand, the life time from $v$ to $v-1$
transition is usually very long. A larger decay rate $\gamma$ is
preferred so that the transition can be observed before other processes
such as chemical reaction destroy the state. For this reason, a stronger
dipole moment is preferred. Table \ref{tab:polmol} summarizes the
relevant parameters for different molecules which are common for cold
molecule experiments. Typically the total number of molecules is about
a few thousands, being confined within a trap with a diameter $d\approx10\sim100\mu$m.
We also lists the values of OD corresponding to currently realizable
experiments, showing that quite a range of different molecules should
be expected to have superradiance. (From this table, one can see that
LiCs is not a good examplifying system because of its light weight
and relatively poor OD while KCs and RbCs are best candidates.)

Vibrational states form a multilevel ladder where the low-lying levels
are approximately harmonic. Approaching the continuum threshold, the
level spacing is getting narrower when the potential curve becomes
more anharmonic. For a typical molecule like LiCs, the number of quasi-harmonic
vibrational levels can be up to a few tens near the potential minimum.
In experiments where the molecules only have a few vibrational excitations,
this uniform portion of the vibration ladder should be the main stage
where spectroscopy is performed. That is, most of the vibrational
transitions are (nearly) degenerate. In such a situation we are urged
to consider multilevel effects because the virtual photon exchange,
which is the main mechanism to build up the Dicke coherence, can now
take place correlating different levels.

\begin{figure}
\begin{centering}
\includegraphics[width=8cm]{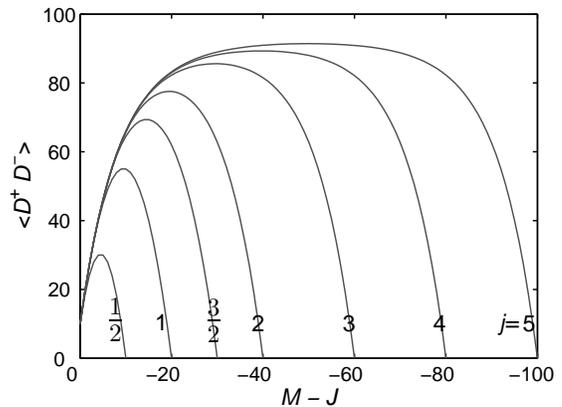}
\par\end{centering}

\caption{\label{fig:ddm}The enhancement factor $\langle D^{+}D^{-}\rangle$
versus $M-J$ in the Dicke picture. Note that the left- (right-) hand
side corresponds to upper (lower) levels. Different curves correspond
to different spin-$j$ cases. Here we have $N=10$ and $J=Nj$.}
\end{figure}

In order to see the multilevel effect we reexamine the enhancement
factor $\langle JM|D^{+}D^{-}|JM\rangle$ in the Dicke picture. Note
that in a multilevel case the molecule can be described as a spin-$j$
particle. Therefore $J=Nj$; $-J\le M\le J$; $D^{+}\equiv\sum_{i}D_{i}^{+}$
with $i$th particle $D_{i}^{+}\equiv\sum_{m=-j}^{j-1}|m+1\rangle_{i}\langle m|$.
The generalized Dicke states for spin-$j$ is given by
\begin{equation}
|J,M\rangle=\sqrt{\frac{(J+M)!}{(2J)!(J-M)!}}(\hat{J}^{-})^{J-M}|J,J\rangle,
\end{equation}
where $\hat{J}^{-}\equiv\sum_{i}\hat{j}_{i}^{-}$, which satisfy
\begin{equation}
\begin{aligned}\hat{J}^{\pm}|J,M\rangle & =\sqrt{J(J+1)-M(M\pm1)}|J,M\pm1\rangle\\
\hat{j}_{i}^{\pm}|j,m\rangle & =\sqrt{j(j+1)-m(m\pm1)}|j,m\pm1\rangle_{i}
\end{aligned}
\end{equation}
with $m$ indexing the single particle vibrational level. Further
calculation yields \cite{Lin2012,lin2012adv6} the enhancement factor
$\langle D^{+}D^{-}\rangle_{JM}=N\langle D_{i}^{+}D_{i}^{-}\rangle+N(N-1)\langle D_{i}^{+}D_{j}^{-}\rangle$
with
\begin{align}
\langle D_{i}^{+}D_{i}^{-}\rangle & =1-\langle j,J-j;-j,M+j|J,M\rangle^{2}
\end{align}
and
\begin{align}
 & \langle D_{i}^{+}D_{j}^{-}\rangle=\sum_{m_{1},m_{2}}\Big[\langle j,J-j;m_{1},M-m_{1}|J,M\rangle\nonumber \\
 & \times\langle j,J-2j;m_{2},M-m_{1}-m_{2}|J-j,M-m_{1}\rangle\nonumber \\
 & \times\langle j,J-j;m_{1}-1,M-m_{1}+1|J,M\rangle\\
 & \times\langle j,J-2j;m_{2}+1,M-m_{1}-m_{2}|J-j,M-m_{1}+1\rangle\Big].\nonumber 
\end{align}

Figure \ref{fig:ddm} plots the enhancement factor $\langle D^{+}D^{-}\rangle_{JM}$
as a function of $M-J$ for various spin-$j$ particles. It is shown
that the decay rate increases in general as the number of levels increases.
At $M\approx J$, all the curves coincide, indicating that they share
the same emission behavior while starting emitting. The enhancement
continues to grow as $M$ lowers before it peaks around $M\approx0$;
then it turns to decrease with descending $M$ and finally vanishes
at $M=-J$. It is worth noting that the curve forms a plateau for
large $j$ around a range of middle $M$, and most importantly the
ascending part of the curve keeps invariant as $j$ increases. This
suggests that the exact number of levels is not important in determining
the initial emission behavior as long as the number of levels is large
enough. This is valid in typical vibrational spectroscopy within a
small timescale (compared to $\gamma^{-1}$).

\section{Two-body master equation\label{sec:Formalism}}

The previous approach concerning Dicke superradiance in the multilevel
case does not take into account the finite spatial size of the sample
and other decoherence mechanism such as dephasing. These additional
factors can be contained by considering explicitly the actual dipole-dipole
interaction. On one hand, dipole-dipole interaction induced spin (two-level
particle) flip-flop, or the so-called virtual photon exchange, is
the major mechanism for generating and maintaining the Dicke type
of coherence. On the other hand, the dipole-dipole interaction ($\sim r^{-3}$,
where $r$ is the inter-particle distance) also introduces dephasing
that suppresses superradiance. These effects can be contained in the
microscopic Hamiltonian \cite{lin2012adv6}:
\begin{align}
H & =\underbrace{H_{\text{atom}}+H_{\text{field}}-\sum_{i\notin\{1,2\}}\mathbf{p}_{i}\cdot\mathbf{E}_{i}(t)}_{H_{0}}\;\underbrace{-\sum_{i=1}^{2}\mathbf{p}_{i}\cdot\mathbf{E}_{i}(t))}_{V},\label{eq:Ham_origin}
\end{align}
where $V$ represents two probe particles ($i=1,2$) interacting with
the local field due to environmental particles as well as the external
field, whose degrees of freedom have been absorbed in $H_{0}$. In
this approach, $V$ is considered to be a perturbative term with respect
to $H_{0}$. Our goal is to derive an effective description for the
two probe atoms in a mean-field manner, which can be thought as the
next order correction beyond the single particle mean-field approximation.
Our strategy is to integrate out irrelevant degrees of freedom such
as field variables and environmental dipoles by means of the Schwinger-Keldysh
Green's function approach. Here a Gaussian field is assumed so that
we are allowed to only keep up to the second order of field correlation.
Then the induced rate is determined by the field-field correlation.
The detail of derivation can be found in \cite{Lin2012,lin2012adv6}
and we arrive at the effective master equation: 
\begin{align}
\dot{\rho}= & -\frac{\Gamma}{2}\sum_{i=1}^{2}\big(\rho D_{i}^{-}D_{i}^{+}+D_{i}^{-}D_{i}^{+}\rho-2D_{i}^{+}\rho D_{i}^{-}\big)\nonumber \\
 & -\frac{\Gamma+\gamma}{2}\sum_{i=1}^{2}\big(\rho D_{i}^{+}D_{i}^{-}+D_{i}^{+}D_{i}^{-}\rho-2D_{i}^{-}\rho D_{i}^{+}\big)\label{eq:mainmaster}\\
 & -\bar{\Gamma}\sum_{i\neq j}\big(\rho D_{i}^{-}D_{j}^{+}+D_{i}^{-}D_{j}^{+}\rho-2D_{j}^{+}\rho D_{i}^{-}\big),\nonumber 
\end{align}
where $\rho$ is the two-body density matrix whose dimension is $(2j+1)\times(2j+1)$,
and 
\begin{align}
\Gamma & =\gamma(e^{2\zeta}-1)\frac{A(t)}{V(t)}+2\mathcal{C}^{2}\varrho^{4}\frac{\gamma^{2}I(\zeta,\varrho)}{\Gamma+\gamma/2}Y(t)\\
\bar{\Gamma} & =\frac{\gamma^{2}I(\zeta,\varrho)}{\Gamma+\gamma/2}\left[3\mathcal{C}\varrho A(t)+2\mathcal{C}^{2}\varrho^{4}Y(t)\right]
\end{align}
with
\begin{align}
A(t) & =\sum_{m=-j+1}^{j}\rho_{mm}^{(1)}\nonumber \\
V(t) & =\rho_{jj}^{(1)}-\rho_{-j,-j}^{(1)}\\
Y(t) & =\sum_{m,m^{\prime}=-j}^{j-1}\rho_{m+1,m;m^{\prime},m^{\prime}+1}.\nonumber 
\end{align}
Here $\rho^{(1)}\equiv\frac{1}{2}\sum_{i=1,2}\text{tr}_{i}[\rho]$
denotes the reduced single-particle density matrix and $\rho_{ab;cd}\equiv\frac{1}{2}[\langle a,c|\rho|b,d\rangle+\langle c,a|\rho|d,b\rangle]$,
where by $|a,c\rangle$ we denote $|a\rangle_{1}\otimes|c\rangle_{2}$.
For both the factor $\frac{1}{2}$ comes from averaging the interchanging
of two particles. Note that interchange symmetry requires $\rho_{ab;cd}=\rho_{cd,ab}$
and $\rho_{ab;cd}^{\ast}=\rho_{ba;dc}$. The cooperativity parameter
is defined as $\mathcal{C}\equiv2\pi\mathcal{N}c^{3}/\omega_{0}^{3}$
with the particle density $\mathcal{N}$; $\varrho\equiv\omega_{0}d/(2c)$
characterizes the system size $d$ in units of the radiation wavelength.
Function $I(\zeta,\varrho)\equiv e^{2\zeta}[(\zeta-1)^{2}+\varrho^{2}]/(\zeta^{2}+\varrho^{2})^{2}$.
With these definitions, $\text{OD}=4\pi\mathcal{C}\varrho$.

Each term of master equation (\ref{eq:mainmaster}) can be understood
as follows: The first line accounts for population pumping with the
induced rate $\Gamma$. The opposite process, i.e., decay, is also
characterized by $\Gamma$, as presented in the second line. In addition,
the free-space spontaneous decay with a rate $\gamma$ takes place
and is contained in the second line. Note that at a finite temperature,
the thermal contribution of blackbody radiation may cause extra incoherent
pumping and decay. This effect can be introduced by adding an additional
rate $B\approx\gamma/(e^{\beta\hbar\omega_{0}}-1)$ with $\beta^{-1}=k_{B}T$
\cite{deiglmayr2011epjd} to $\Gamma$. Further, the population pumping
corresponds to photon re-absorption (real photons only, not virtual
photons). This, however, may depends on the scattering rate of photons
and particles. In order to characterize the efficiency of re-absorption,
we replace in the first line $\Gamma\longrightarrow s\Gamma$, where
the factor $s=1$ corresponds to maximal re-absorption of emitted
photons and $s=0$ represents loss of emitted photons entirely. Although
at this point we do not estimate $s$ quantitatively, these two extremes
$s=0,1$ should be able to give the lower and upper bounds for the
actual behavior. The third line corresponds to generation and reduction
of inter-particle coherence with a rate $\bar{\Gamma}$. Finally,
considering the current experimental setup, where there is inevitably
particle loss out of the trap, one more term $-\gamma_{L}\rho$ must
be added into to master equation (\ref{eq:mainmaster}). Note that
this term does not preserve unity trace of the density matrix as time
evolves.

\section{Results and discussion\label{sec:Result}}

\subsection{Multilevel enhancement}

\begin{figure}
\begin{centering}
\includegraphics[width=8cm]{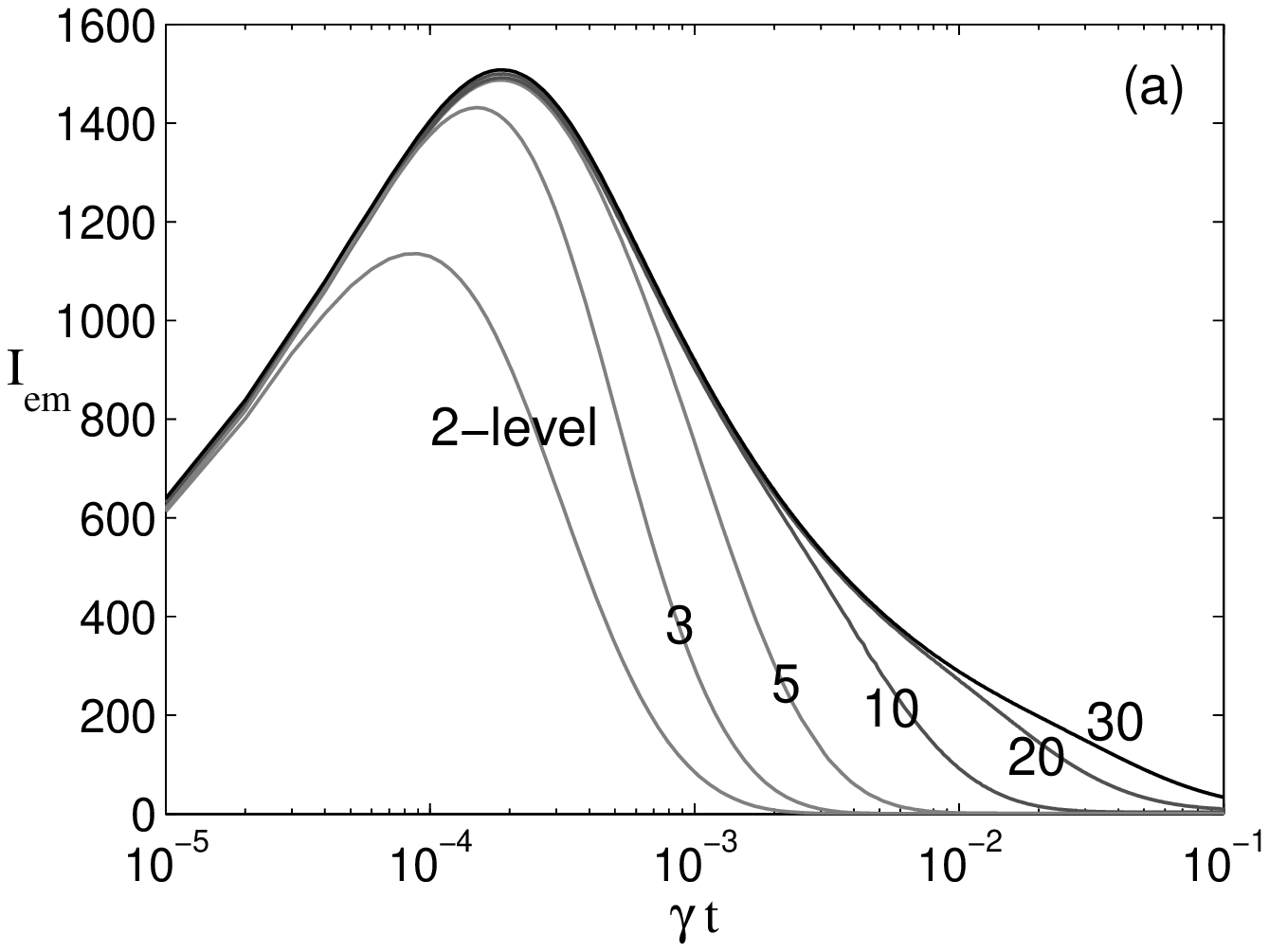}
\par\end{centering}

\begin{centering}
\includegraphics[width=8cm]{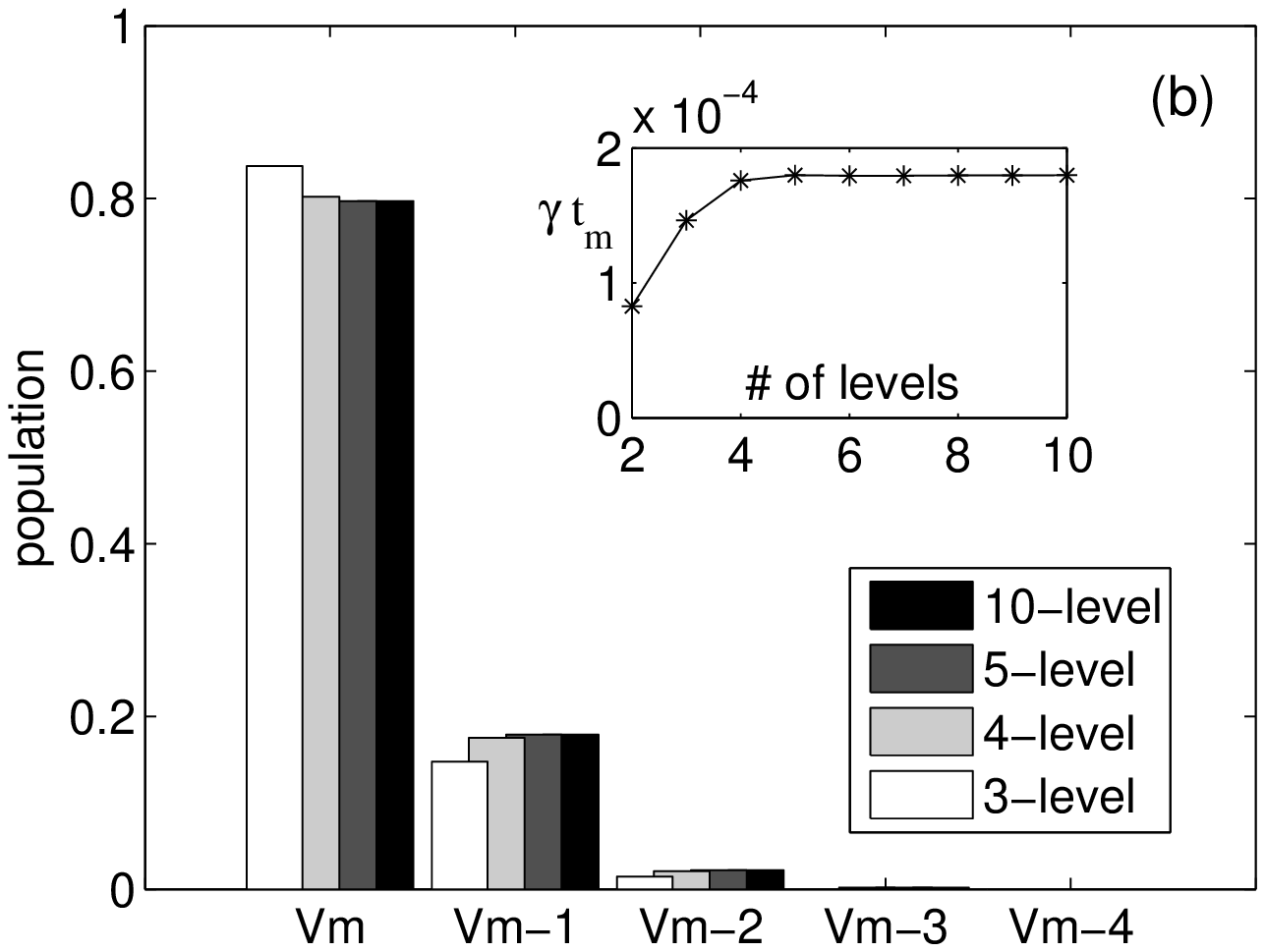}
\par\end{centering}

\caption{\label{fig:multilevel} (a) Emission intensity as a function of time.
Different curves correspond to different number of vibrational levels.
For all data, $\text{OD}=3.8\times10^{4}$. (b) Population distribution
of the five highest levels for 3, 4, 5, and 10-level cases at peak
intensity (when $t=t_{m}$). $V_{m}$ denotes the highest level where
the whole system is initially populated. Inset: $\gamma t_{m}$ as
a function of the number of levels.}
\end{figure}

We have investigated the multilevel superradiance in great detail
in \cite{Lin2012}. For completeness, here we include a brief discussion
for this effect. Such an enhancement is due to the fact that the cross-level
virtual photon exchange is more possible. In a two-level system, when
a particle emits a real photon but fails to keep correlation to other
particles due to certain dephasing, it no longer participates in the
superradiance coherence unless it gets excited later and re-enters
the game. This is different in the multilevel case because a highly
excited particle can be involved in manybody correlation for several
times of its transitions before the ground state is reached. Fig.
\ref{fig:multilevel} (a) shows the emission intensity $I_{\text{em}}$
as a function of time, where $I_{\text{em}}\equiv\hbar\omega_{0}\frac{d}{dt}\sum_{v}v\rho_{vv}^{(1)}$
corresponding to the overall outgoing radiated energy per unit time.
These curves are dramatically different from an exponential profile
yielded by single atoms. The initial increase and hump in the intensity
is the signature for superradiance. Note that the rising parts of
the curves coincide for various level numbers, except that a larger
discrepancy is observed in the two-level case. This coincidence even
extends for longer times in the many level cases; the hump is bounded
in the limit of large level numbers. This can be understood by looking
at Fig. \ref{fig:multilevel} (b), where we show the top five levels'
population distribution (diagonal terms of the reduced single particle
density matrix) over the vibrational ladder at the time when the intensity
value is peaked. We can see that the distribution is almost identical
for various cases. This happens because the initial emission behavior
is determined only by the nearby levels. Those levels far below do
not contribute.

\subsection{Comparison with experiment}

\begin{figure}
\begin{centering}
\includegraphics[width=8.5cm]{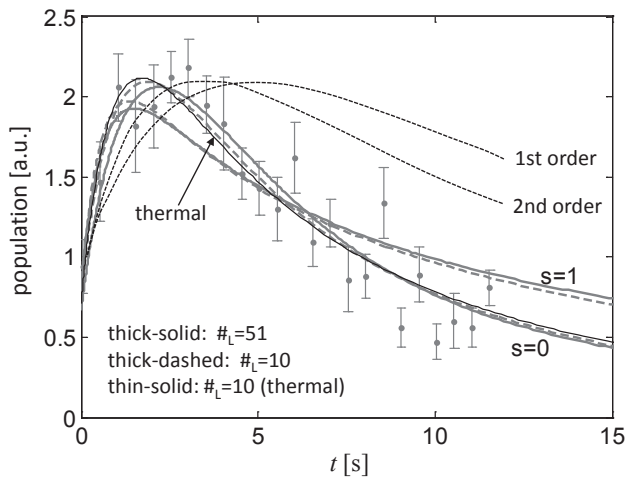}
\par\end{centering}

\caption{\label{fig:v3curve}Fitting curves by adjusting the ratio $N_{4}/N_{3}$
between the population of levels $v=3$ and $v=4$. For $s=1$, i.e.,
the re-absorption is considered, the optimized ratio $N_{4}/N_{3}=7.47$;
for $s=0$, i.e. the re-absorption is neglected, the ratio $N_{4}/N_{3}=6.72$.
An optimization is also taken by overall rescaling the $y$ value
for best fit. For comparison, two simulated curves from \cite{deiglmayr2011epjd}
are plotted, corresponding to inclusion of the first-order and second-order
particle losses, respectively. The thin-solid line corresponds to
the case when the balckbody radiation is considered, with the optimized
ratio $N_{4}/N_{3}=7.55$. The resultant difference, however, is rather
small. The scattered dots with error bars are the experimental data.}
\end{figure}

In usual cold molecule experiments, individual atoms are first loaded
and cooled in a trap and then through photoassociation Feshbach molecules
can be formed. In most situations, the electronic state of the molecules
is kept on the ground state but not for vibrational and rotational
states. In a recent experiment done by Deiglmayr\textit{ et al.,}
the vibrational population redistribution of LiCs molecules was investigated
\cite{deiglmayr2011epjd}. An unusual discrepancy was found between
the experimental data and the simulated curves given by rate equation
approaches based on Einstein's A-coefficients and inclusion of blackbody
radiation. Here we apply the superradiance formalism to their experimental
setup, where $\text{OD}\approx500$. In Fig. \ref{fig:v3curve} we
show the time evolution of the population profile corresponding to
level $v=3$, directly comparable to the photoassociation measurement
outcome.

We fit the experimental data as follows: From \cite{deiglmayr2011epjd}
we find the ratios of the lowest ten vibrational levels. In the experiment,
however, most electrons ($\sim$70\%) are still in some of the upper
levels, with unknown distribution. Thus, we have to allow for at least
one fitting parameter to take this unknown into account. In our case,
for ease of calculation, this fitting parameter is the ratio between
the population of levels $v=3$ and $v=4$. The curves are then calculated
for $51$ total levels and an approximation where only the lowest
$10$ of those $51$ levels are used (Fig. \ref{fig:v3curve} shows
that there is nearly no difference between those two cases.)

In Fig. \ref{fig:v3curve}, the scattered dots with error bars represent
the experimental data and the thick-solid (thick-dashed) lines correspond
to the overall 51-level (10-level truncated) case. For comparison,
we plot $s=0$ and $s=1$ cases corresponding to the upper and lower
bounds, respectively, when re-absorption efficiency is $0$ or maximized.
We find that the $s=0$ curves show a very good agreement with the
experimental data, suggesting that re-absorption is not a significant
effect when the trap is transparent to radiation. Most importantly,
such agreement implies that superradiance (or at least the inter-particle
cooperativity) plays an important role in determining the relaxation
behavior. It is notable that, in this experiment, $\text{OD}\approx500$
is very low, even lower than the values suggested in Table \ref{tab:polmol}.
It is actually near the boundary of superradiance and ASE. Note that
ASE as well as superradiance are both caused by the dipole-dipole
interaction. But in ASE the decoherence dominates and therefore the
inter-particle coherence is less pronounced. It should be emphasized
that the ASE mechanism is also contained under the framework of our
approach. In fact, LiCs is a rather bad choice for demonstrating superradiance
because of its low reduced mass. A heavier species such as RbCs under
similar experimental setup will do better (see Table \ref{tab:polmol}).

\subsection{Blackbody radiation}

An additional curve (thin-solid line) is also plotted in Fig. \ref{fig:v3curve},
taking into account the thermal contribution of blackbody radiation.
But the deviation is rather small. Here we briefly discuss the role
of the blackbody radiation repumping and decay. In Eq. (\ref{eq:mainmaster})
it serves as a factor in addition to $\Gamma$. At room temperature
$T=300$K, we can estimate $B=0.75\gamma$. This is usually a significant
quantity in the single particle case but not in the superradiance
case, where $\Gamma$ is typically 2 orders of magnitude larger than
$\gamma$ for $\text{OD}=10^{2}\sim10^{3}$.

\section{Conclusion\label{sec:Conclusion}}

The vibrational relaxation behavior of cold molecules displays superradiant
behavior when their density is large so their particle cooperativity
must be considered. We have developed a mean-field method that explicitly
take into account the actual dipole contribution as well as the multilevel
effect. Our theory works rather well describing the experimental data
from measuring the relaxation of vibrational population for cold molecules.
This indicates that the cooperativity has a significant effect regarding
spectroscopy in most ultracold molecules experiments, including vibrational
and/or rotational transitions. Our calculations suggest that considerable
cooperative effects can be expected from a large amount of ultracold
molecule experiments if vibrational and rotational transitions are
considered. In general, for all setups with an OD density of the order
of $10^{2}\sim10^{3}$ or higher, superradiance will play an important
role.

Some open questions remain such as how superradiance is modified by
(1) the anharmonicity of the vibrational ladder, (2) the non-constant
transition dipole moments from level to level, and (3) the trap-induced
inhomogeneity in density. They are, however, at this point not expected
to have a major qualitative impact on the results.
\begin{acknowledgments}
We would like to thank NSF and AFOSR for funding, and M. Repp, J.
Deiglmayr, and M. Weidem\"uller for discussions.\end{acknowledgments}

\end{document}